\documentclass[amssymb,amsmath,aps,prl,twocolumn,superscriptaddress,showpacs]{revtex4}
\usepackage[dvips]{graphicx}
\usepackage{dcolumn}
\usepackage{bm}
\usepackage{fancyhdr}
\usepackage{float}
\usepackage{ifthen}
\usepackage{eurosym}
\usepackage{color}

\begin{document}
\preprint{NaxCoO2}
\title{1D to 2D Na$^+$ Ion Diffusion Inherently Linked to Structural Transitions in Na$_{0.7}$CoO$_{2}$}
\vspace{-7mm}
\author{M.~Medarde}
  \email{marisa.medarde@psi.ch}
\affiliation{Laboratory for Developments and Methods, Paul Scherrer Institut, CH-5232 Villigen PSI, Switzerland}
\author{M.~Mena}
\author{J.~L.~Gavilano}
\affiliation{Laboratory for Neutron Scattering, Paul Scherrer Institut, CH-5232 Villigen PSI, Switzerland}
\author{E.~Pomjakushina}
\affiliation{Laboratory for Developments and Methods, Paul Scherrer Institut, CH-5232 Villigen PSI, Switzerland}
\author{J.~Sugiyama}
\affiliation{Toyota Central Research and Development Labs. Inc., Nagakute, Aichi 480-1192, Japan}
\author{K.~Kamazawa}
\affiliation{Comprehensive Research Organization for Science and Society (CROSS), Tokai, Ibaragi 319-1106, Japan}
\author{V.~Yu.~Pomjakushin}
\author{D.~Sheptyakov}
\affiliation{Laboratory for Neutron Scattering, Paul Scherrer Institut, CH-5232 Villigen PSI, Switzerland}
\author{B.~Batlogg}
\author{H.~R.~Ott}
\affiliation{Laboratory for Solid state physics, ETH Z\"{u}rich, CH-8093 Z\"{u}rich, Switzerland}
\author{M.~M\aa{}nsson}
\affiliation{Laboratory for Solid state physics, ETH Z\"{u}rich, CH-8093 Z\"{u}rich, Switzerland}
\affiliation{Laboratory for Neutron Scattering, Paul Scherrer Institut, CH-5232 Villigen PSI, Switzerland}
\author{F.~Juranyi}
\affiliation{Laboratory for Neutron Scattering, Paul Scherrer Institut, CH-5232 Villigen PSI, Switzerland}


\date{\today}

\begin{abstract}
We report the observation of a stepwise ``melting'' of the low-temperature Na-vacancy order in the layered transition metal oxide Na$_{0.7}$CoO$_{2}$. High-resolution neutron powder diffraction indicates the existence of two first-order structural transitions, one at $T_{1}~\approx~290$~K, followed by a second at $T_{2}~\approx~400$~K. Detailed analysis strongly suggests that both transitions are linked to changes in the Na mobility. Our data are consistent with a two-step disappearance of Na-vacancy order through the successive opening of first quasi-1D ($T_{1}$ $>$ $T$ $>$ $T_{2}$) and then 2D ($T$ $>$ $T_{2}$) Na diffusion paths. These results shed new light on previous, seemingly incompatible, experimental interpretations regarding the relationship between Na-vacancy order and Na dynamics in this material. They also represent an important step towards the tuning of physical properties and the design of tailored functional materials through an improved control and understanding of ionic diffusion.
\end{abstract}

\pacs{71.27.+a, 71.30.+h, 71.45.Lr, 61.05.fm}
\maketitle

Amongst correlated electron systems, layered transition metal oxides (TMO) have attracted a particular interest in condensed matter physics for their intriguing magnetic \cite{Anderson_73} and electronic \cite{Bednorz_86} properties. The key to their understanding is believed to lie in the fundamental physics of the TMO planes, with the structural ion-units sandwiched between them acting merely as passive charge reservoirs. In parallel, many of these materials are in the center of attention for applied sciences, in particular in the field of rechargeable batteries \cite{Linden_10}. Here the focus is instead on the intermediary ionic layers and in particular their \emph{dynamic} properties, while the role of TMO planes is considered as secondary. Very recently these two fields have been unified under the framework of the layered Na$_{x}$CoO$_{2}$ oxyde family \cite{Takada_03}. While its magnetic \cite{Sugiyama_04} and electronic \cite{Balicas_08} properties change dramatically with the number of holes/Co (i.e. $x$) in the Co-O layers \cite{Delmas_81}, it has lately become evident that also the potential landscape created by ordered Na vacancies \cite{Roger_07,Julien_08} as well as the Na dynamics \cite{Schulze_09,Weller_09} play equally important roles. This has hinted at the possibility of tuning  physical properties \cite{Julien_08} through the control of the self-assembling ion-texture (i.e. periodic potential landscape) on a sub-nanoscale. The \emph{key-issue} is here to understand the solid state dynamics at the atomic level.

\vspace{-1mm}
Below we present a high-resolution neutron powder diffraction (NPD) study which clearly reveals how Na dynamics and changes in the crystal structure are inherently linked in the Na$_{x}$CoO$_{2}$ system. For this purpose we have chosen Na$_{0.7}$CoO$_{2}$, whose Na-vacancy ordered array is amongst the most stable of the Na$_{x}$CoO$_{2}$ family \cite{Shu_08}. The symmetry of this pattern has been investigated in the past \cite{Mukhamedshin_04,Roger_07,Morris_09,FTHuang_09}, but the actual periodicity as well as the size and arrangement of Na-vacancy clusters are still subject of debate. The temperature dependence of this array has been much less investigated, but a few groups have predicted \cite{Hinuma_08} as well as reported \cite{Weller_09,Morris_09,QHuang_04} the existence of a first order transition at $T_{1}$ $\sim$ 290 K. The nature of the high temperature phase ($T>T_{1}$) is so far unclear. Single-crystal neutron diffraction studies suggest the stabilization of a static stripe-like ordered Na-vacancy array \cite{Morris_09}. In contrast,  NMR \cite{Gavilano_04,Weller_09} and $\mu^{+}$SR investigations \cite{Schulze_09} indicate an increasing Na diffusion for $T>200$~K and a ``melting' of the low-temperature Na-vacancy array above $T_{1}$.

With the aim of clarifying these two apparently incompatible results we have re-investigated the crystal structure of Na$_{0.7}$CoO$_{2}$ as a function of temperature. Details about the synthesis of the powder sample used in this study as described in the Supplemental Material [SM]. A series of NPD patterns ($T=10-475$~K) were recorded at the high resolution diffractometer HRPT \cite{HRPT} (monochromator: Ge (533), $\lambda = 1.494(1)~\text{\AA}$) of the Swiss Neutron Spallation Source SINQ, Paul Scherrer Institut, PSI Villigen, Switzerland. Scattering from the sample environment (cryo-furnace) was suppressed using an oscillating radial collimator. Analysis of the data, including the explicit consideration of absorption effects, was done using the Rietveld refinement package \textsf{FullProf Suite} \cite{Fullprof}[SM].

The Na$_{x}$CoO$_{2}$ family is characterized by a layered structure with alternating CoO$_{2}$ slabs and interstitial, partially occupied, Na$_{x}$ layers [SM]. Two crystallographic, non-equivalent Na sites are located at the nodes of two interpenetrating triangular lattices: Na1, which lies above and below the Co sites in the neighboring CoO$_2$ layers, and Na2, located at the center of a Co trigonal prism. The average crystal structure of Na$_{0.7}$CoO$_{2}$ has been described using a primitive unit cell of $P6_{3}/mmc$ symmetry (\textbf{a}$_{h}\approx2.83$~{\AA}, \textbf{c}$_{h}\approx10.88$~{\AA} [SM]). Extra superstructure reflections have been observed above and below $T_{1}$ and attributed to different Na-vacancy ordered patterns. The corresponding supercells have been reported to be commensurate with the underlying $P6_{3}/mmc$ lattice which, according to \cite{Roger_07, Morris_09,FTHuang_09}, remains metrically hexagonal for $T=1.5-350$~K.

\begin{figure}[tbh]
\includegraphics[keepaspectratio=true,width=75 mm]{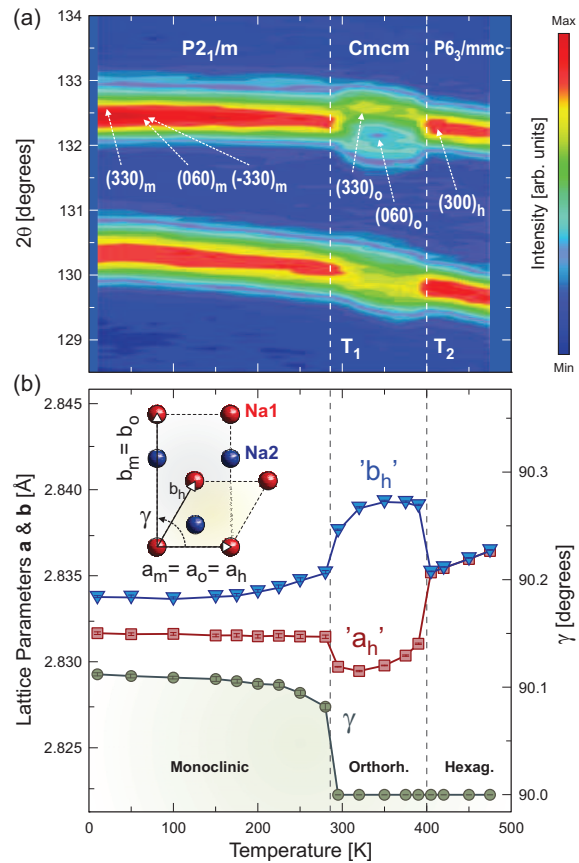}
\vspace{-3mm}
\caption{(Color online)(a) 2D contour plot showing the temperature dependence of the neutron powder diffraction patterns for Na$_{0.7}$CoO$_{2}$. (b) Relationship between the in-plane hexagonal (\textbf{h}), orthorhombic (\textbf{o}) and monoclinic (\textbf{m}) lattice parameters together with its thermal evolution. The vertical lines indicate the two transition temperatures $T_{1}~\approx~290$~K and $T_{2}~\approx~400$~K.}
\label{fig:Data}\vspace{-5mm}
\end{figure}

The present data reveal new details concerning the metrics of the average unit cell. The temperature dependence of the high-angle region in the neutron diffraction patterns [Fig.~\ref{fig:Data}(a)] clearly shows the presence of two first-order structural phase transitions involving changes in the P6$_{3}$/mmc fundamental reflections. The first of them occurs at $T_{1}~\approx~290$~K, coinciding with the anomaly described in previous studies \cite{Morris_09,QHuang_04,Weller_09}. A second transition, not previously reported, is found at higher temperatures ($T_{2}~\approx~400$~K). The neutron diffraction patterns above $T_{2}$ can indeed be refined using the $P6_{3}/mmc$ space group [SM]. This is however not possible for the intermediate region between  $T_{1}$ and $T_{2}$, where a splitting of the hexagonal reflections with nonzero Miller indices \textit{h} and/or \textit{k} is clearly visible [see Fig.~\ref{fig:Data}(a)]. This indicates that the \textbf{a$_{h}$} and  \textbf{b$_{h}$} hexagonal axis are inequivalent within this temperature range. The diffraction patterns can be indexed on an orthorhombic cell  \textbf{a$_{o}$}~=~\textbf{a$_{h}$},  \textbf{b$_{o}$}~=~\textbf{b$_{h}\cdot\sqrt{3}$}, \textbf{c$_{o}$}~=~\textbf{c$_{h}$} (\textbf{a$_{h}$} and \textbf{b$_{h}$} denote here pseudohexagonal axes) and refined using the \textit{Cmcm} space group, which is one of the maximal non-isomorphic subgroups of P6$_{3}$/mmc. The relationship between the hexagonal and orthorhombic cells is schematically shown in the inset of Fig.~\ref{fig:Data}(b). Fit details and tables with the atomic positions are given in [SM].

The splitting of the \textbf{a$_{h}$} and \textbf{b$_{h}$} orthorhombic axes at $T_2$ upon cooling gives rise to an expansion of the \textbf{ab}-plane, which coincides with a much weaker increase of \textbf{c} [SM].  Below $T_{1}$ it decreases again, although it remains nonzero at the lowest investigated temperature (Fig.~\ref{fig:Data}(b)). The Rietveld fits using either the \textit{P6$_{3}$/mmc} or \textit{Cmcm} space groups are significantly worse in this temperature region. In particular, the extra splitting/broadening of some of the high-angle reflections cannot be reproduced without decreasing the symmetry from orthorhombic to monoclinic [SM]. The fits using the \textit{P2$_{1}$/m} space group and a unit cell very similar to that of \textit{Cmcm} with $\gamma$ $>$ 90$^{\circ}$ (90.117(2)$^{\circ}$ at 10K) result in a clear improvement (see fit details on [SM]). However, the agreement between the observed and calculated intensities is worse than in the intermediate and high temperature regions (see difference patterns and $\chi$$^{2}$ values in [SM]). This indicates the existence of subtle structural features which can not be reproduced by the average \textit{P2$_{1}$/m} monoclinic cell. Such behavior is consistent with the existence of complex Na-vacancy ordering below $T_{1}$.

\begin{figure*}[tbh]
\includegraphics[keepaspectratio=true,width=100 mm]{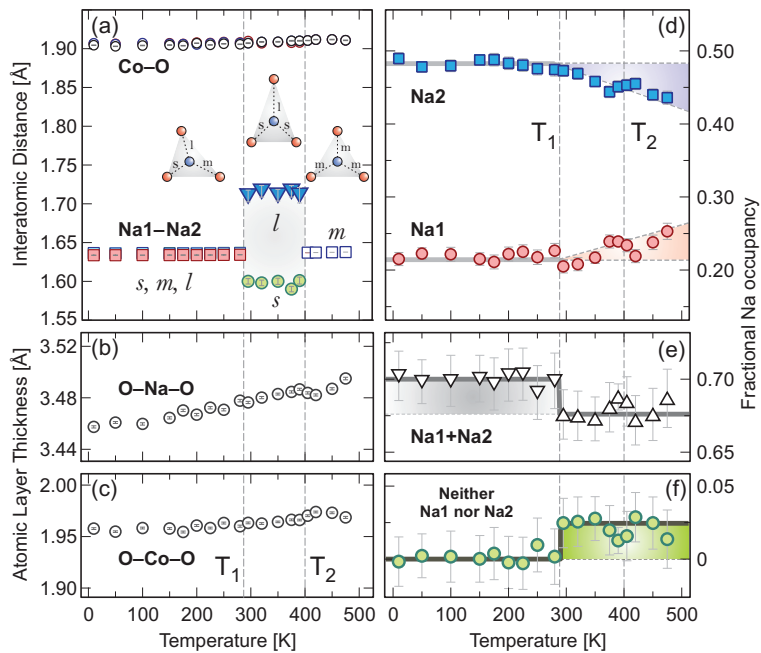}
\vspace{-3mm}
\caption{(Color online) Temperature dependence of (a) the Na1-Na2 and Co-O distances, (b) the thickness of the O-Na-O layer and (c) that of the CoO$_2$ slabs. (d-f) Na1 and Na2 site occupancies. A value of 1 corresponds to a site fully occupied. In absence of diffusion $n_{Na1}$ + $n_{Na2}$ corresponds to the total Na content ($x$ in Na$_{x}$CoO$_{2}$). Inset in (a) is a schematic representation of the Na-triangle distortions and Na1-Na2 lengths (\textit{l, m, s}) for the different space groups (see also Fig.1 of [SM]). Solid lines and shadings are guides to the eye.}
\label{fig:Fig2}\vspace{-3mm}
\end{figure*}

We shall show now that the occurrence of this "quasi-reentrant" structural rearrangement lies in the connection between structural changes and Na dynamics in Na$_{0.7}$CoO$_{2}$. The evolution of the in-plane distances between neighboring Na$^+$ ions is shown in Fig.~\ref{fig:Fig2}(a). The average low-temperature monoclinic structure used to refine our data allows three distinct, although nearly indistinguishable Na1-Na2 distances \textit{l} (long), \textit{m} (medium) and \textit{s} (short), see the insets of Fig.~\ref{fig:Fig2}(a) and [SM]. Because this structural model does not take into account the Na-vacancy order, the actual Na1-Na2 distances may be differ from these values. However, the deviations are expected to be smaller than 0.2 $\text{\AA}$)(see [SM]). By entering the orthorhombic phase 2 of these 3 distances become identical and shorter (-2.5\%) whereas the third one undergoes a dramatic increase (+5.2\%). The difference between the long (\textit{l}) and the short one (\textit{s}) changes very little between $T_1$ and $T_2$. Then, it drops to zero at the transition to the high temperature hexagonal phase ($T>T_2$), where the three Na1-Na2 distances become identical (\textit{m}).

The changes of the Na1-Na2 distances at $T_1$ and $T_2$ are about one order of magnitude larger than those undergone by any other interatomic distances in the structure. This can be appreciated in Fig.~\ref{fig:Fig2}(a), where the temperature dependence of the three Co-O distances is also displayed.  Particularly noticeable is the absence of significant discontinuities in the \textbf{c} lattice parameter since, according to other studies, displacements of the Co and/or O atoms along this crystallographic axis would be expected as response to the changes in the Na1/Na2 occupancies. This lack of anomalies is not due to compensation effects, as demonstrated by the smooth temperature dependence of the thickness of the CoO$_2$ slabs and the O-Na-O layers [Figs.~\ref{fig:Fig2}(b-c)]. We thus conclude that the main coherent atomic displacements occurring at $T_{1}$ and $T_{2}$ are confined to the Na-ion planes.

The existence of a correlation between structural changes and Na dynamics is first indicated by the thermal evolution of the Na1/Na2 fractional occupancies, $n_{Na1}$ and $n_{Na2}$ which, together with their sum $n_{Na1}$ + $n_{Na2}$ is shown in Fig.~\ref{fig:Fig2}(d-e). At $T=10$~K, the energetically more favorable Na2 position displays an occupancy approximately twice that of Na1, in agreement with previous reports \cite{Zandbergen_04}. The refined value of $n_{Na1}$ + $n_{Na2}$ ($\sim 0.7$) coincides well with the $x$ value derived from the lattice parameter \textbf{c} \cite{Hinuma_08} and it does not change below $T_1$ within experimental error. This is not the case for $n_{Na1}$ and $n_{Na2}$, whose temperature dependence indicate a weak but progressive Na transfer from the Na2 into the Na1 sites above $\sim$T=200K. At $T_1$ a clear anomaly is observed that coincides with a substantial increase of the Na2 $\rightarrow$ Na1 transfer. Such behavior is consistent with the growing diffusion above $T=200$~K and the "melting" of the Na layers at $T_1$ suggested by NMR \cite{Weller_09} and $\mu^{+}$SR studies \cite{Schulze_09}. An interesting detail is that the sum of the Na content at the Na1 and Na2 sites ($n_{Na1}+n_{Na2}$) remains nearly constant below $T_{1}$ . However, it undergoes a small but significant decrease ($\sim-0.02$) at $T_1$ [Fig.~\ref{fig:Fig2}(e)]. This means that, for $T>T_1$ and within the interaction time of thermal neutrons, approximately 3$\%$ of all the Na$^+$ ions are located away from the Na1 and/or Na2 positions [Fig.~\ref{fig:Fig2}(f)].

Further experimental evidence suggesting an active role of Na dynamics is provided by the temperature dependence of the Debye-Waller (DW) factors. These quantities reflect the reduction of Bragg integrated intensities due to the displacements of atoms from their equilibrium positions. For a harmonic crystal, they can be interpreted in terms of time-averaged mean-square displacements $<$$u^2$$>$ resulting from all normal modes of vibration. However, the $<$$u^2$$>$ values obtained from structural refinements often contain information about other types of displacements, either static or dynamic, which are not included in the structural model.

\begin{figure}[tbh]
\includegraphics[keepaspectratio=true,width=85 mm]{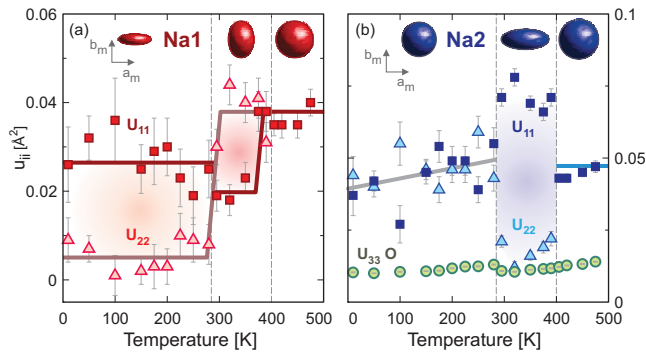}
\vspace{-3mm}
\caption{(Color online) Temperature dependence of the in-plane mean square displacements $u_{11}$ and $u_{22}$ for the Na1 and Na2 sites, as well as the out-of-plane $u_{33}$ for the O site. Top insets display the shape of the projection of the associated Na1/Na2 thermal ellipsoids in the \textbf{ab}-plane}.
\label{fig:Fig3}\vspace{-3mm}
\end{figure}

The evolution of the mean-square displacements $u_{11}$ and $u_{22}$ at the Na1 site along the \textbf{a} and \textbf{b} axes, respectively, (monoclinic/orthorhombic notation) is displayed in Fig.~\ref{fig:Fig3}(a). At $T=10$~K $u_{22}$ is nearly zero, consistent with the reduced thermal motion expected at this temperature. Around $T=200$~K, however, a rapid increase suddenly sets on. In contrast, $u_{11}$ is rather large already at $T=10$~K and displays only a weak $T-$dependence, signaling a static contribution superimposed onto a very weak thermal motion. This is consistent with the existence of Na-vacancy order, proposed in previous studies and suggests a distribution of Na1 positions along the \textbf{a$_{m}$} direction. For the Na2 site, large values of $u_{11}$ and $u_{22}$ below $T_1$ along with a weak $T-$dependence indicate once again the presence of static displacements but, contrary to Na1, the associated distribution of Na2 positions is nearly isotropic (i.e. $u_{11}\approx{}u_{22}$).

\begin{figure*}[tbh]
\includegraphics[keepaspectratio=true,width=175 mm]{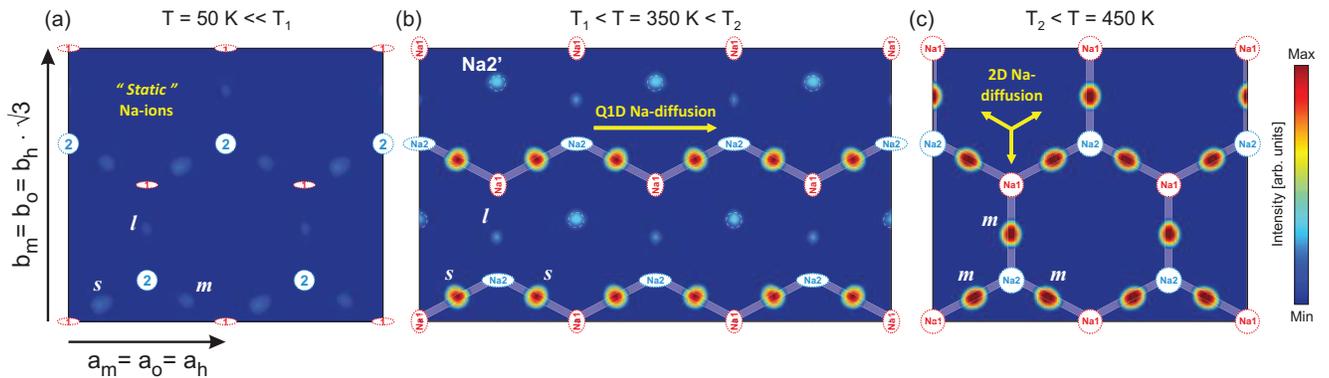}
\vspace{-3mm}
\caption{(Color online) Fourier difference maps of the $z=0.25$ Na planes at $T=50,~320~\rm{and}~450$~K showing the evolution of the residual scattering density in the paths connecting the Na1 and Na2 sites (scale top/bottom: 0.25/0.0875 Fermi). For the three temperature regions there is a clear evolution from (a) Static Na$^+$ ions, to (b) Quasi-1D ion diffusion along the \textbf{a}$_{o}$-axis to finally (c) fully 2D ion diffusion. Thick solid white lines are overlayed together with the Na-ion positions in order to emphasize the diffusion paths. The shape of the Na$^+$ ions has been adapted to the thermal ellipsoids previously shown in Fig.~\ref{fig:Fig3} and the different Na1-Na2 distances (\textit{l, m ,s}) from Fig.~\ref{fig:Fig2}(a) are also indicated.}
\label{fig:Fig4}\vspace{-3mm}
\end{figure*}

At $T_1$ clear anomalies are observed in the Na mean-square displacements $u_{11}$ and $u_{22}$. As shown in Fig.~\ref{fig:Fig3}, the in-plane projection of the Na1 and Na2 thermal ellipsoids drastically change their shape. This may arise either from a new distribution of static displacements, $e.g.$ from a new Na-vacancy arrangement \cite{Morris_09} or from the onset of Na diffusion, as suggested by NMR \cite{Weller_09} and $\mu^{+}$SR measurements \cite{Schulze_09}. Based on the DW factors alone it is not possible to distinguish between these two possibilities. However, the changes in the Na1/Na2 fractional occupancies mentioned above clearly favor the second one. A further arguments is that the largest projection of the Na1/Na2 thermal ellipsoids is now observed along the shortest Na1-Na2 distances ($s$) whereas the contribution along $l$ is significantly smaller [Fig.~\ref{fig:Fig4}b]. The mobility of diffusing ions in ionic conductors is known to largely depend on the connectivity between the available hopping sites. Hence, this observation suggests the activation of quasi-one-dimensional (Q1D) ...Na1$\leftarrow^{s}\rightarrow$Na2$\leftarrow^{s}\rightarrow$Na1$\leftarrow^{s}\rightarrow$Na2... zigzag diffusion pathways at $T_1$ coinciding with the shortening of 2/3 of the Na1-Na2 distances [Fig.~\ref{fig:Fig2}a and Fig.~\ref{fig:Fig4}b]. Above $T_2$ a similar correlation between Na$^+$ motion and structural changes can be established: the thermal ellipsoids of all Na sites become perfectly isotropic coinciding with the equalization of Na1-Na2 distances (\textit{m}) . This suggests a crossover from Q1D to 2D Na$^+$ diffusion at $T_2$.

Further experimental evidence supporting this scenario is given by the Fourier difference maps [SM] which provide information about the residual scattering density contributing to the Bragg reflections not reproduced by neither the structural model. Three representative examples of the results obtained for a Na layer ($z = 0.25$) at $T=50,~350~\rm{and}~450$~K are shown in Fig.~\ref{fig:Fig4} with the Na1-Na2 framework superimposed (see also [SM]). At low temperatures [Fig.~\ref{fig:Fig4}(a)], such Fourier maps are rather featureless and only very low intensity is visible in the paths connecting nearest neighbor Na1-Na2 sites. This suggests that the Na$^+$ ions are virtually static at these temperatures. At $T\approx200$~K the inter-Na intensities slowly start to increase, in line with the anomaly of Na1 $u_{22}$ mean-square displacement as well as previous NMR and $\mu^{+}$SR studies \cite{Weller_09,Schulze_09}. However, it is not until $T_1$ is reached that a very strong increase in the residual scattering density is observed. As seen from Fig.~\ref{fig:Fig4}(b), clear maxima appear between the Na1-Na2 ions along the short $s-$paths, while along the long $l-$path the intensity remains weak. This is in excellent agreement with the picture presented above where Na$^+$ ion diffusion develops through Q1D zigzag channels along the \textbf{a}-axis [see Fig.~\ref{fig:Fig4}(b)]. It should also be noted that there is a small intensity increase at (1/2, 0.85, 0.25) and equivalent positions [labeled Na2' in Fig.~\ref{fig:Fig4}(b)]. This Na site is empty in the low and high temperature phases but becomes partially occupied in the orthorhombic phase due to the displacement of the Na2 position out of center of the triangle formed by the Na1 sites. This displacement results in a small enlargement of the Na2' cavity, which explains the increase in the \textbf{ab}-plane area and unit cell volume for $T_{1}$ $>$ $T$ $>$ $T_{2}$ mentioned in previous sections (see also [SM]). Rietveld refinements carried out including the Na2'-site indicate that it hosts about 2$\%$ of the Na$^+$ ions between $T_1$ and $T_2$, which would correspond to the static part of the 'missing' Na occupancy shown in Fig.~\ref{fig:Fig2}(e-f).

With further increasing temperature, another redistribution of the residual scattering intensity is observed at $T_2$ [see Fig.~\ref{fig:Fig4}(c)]. As the sample enters the hexagonal phase and all Na1-Na2 paths become equivalent ($m$), the previously closed channel along the \textbf{b}$_{m/o}-$axis opens up and the Na$^+$ ion diffusion becomes isotropic within the \textbf{ab}-plane. Such Q1D-to-2D evolution of the Na diffusion mechanism is fully consistent with the conclusions derived from temperature dependence of the interatomic distances and anisotropic DW factors. Ergo, for this compound, structural transitions and Na dynamics are strongly interrelated. Although this behavior is found in other ionic conductors, the present case is particular because Na diffusion not only starts along the shorter pathways: it also seems to trigger the structural transition which creates them, or viceversa. As demonstrated by our analysis, the activation of Na diffusion in a perfectly hexagonal lattice requires relatively high temperatures ($T_2\approx400$~K). Hence, the occurrence of a very small structural distortion enables Q1D diffusion paths to open and Na$^+$ ion diffusion may set in at much lower temperatures ($T_1\approx290$~K).

These findings are not only crucial for the understanding of the Na$_{0.7}$CoO$_{2}$ compound, but also indicate that other materials from the same family may likely share a similar precursor to the 2D diffusive state. Consequently, it should be possible to change the onset temperature for diffusion by applying external perturbations ($e.g.$ pressure) able to enhance the material's own intrinsic driving force to create active diffusion channels. This offers completely new opportunities for an optimized tuning of functional materials in applications such a solid electrolytes or electrodes for novel rechargeable batteries.

We finally address the impact of Na$^+$ ion diffusion on the CoO$_2$ layers. As mentioned previously, no significant anomalies are observed in the Co-O distances. However, the O out-of-plane mean-square displacement $u_{33}$ undergoes a small but sharp decrease at $T_1$ [see Fig.~\ref{fig:Fig3}(e)]. This indicates that changes in Na landscape and the simultaneous onset of Q1D Na diffusion indeed modify the width of the Co-O distance distribution along the \textbf{c}-axis even if the mean value remains unchanged. Such behavior may arise from a modification of a static Na-vacancy order but more likely it is an effect related to the onset of Na$^+$ diffusion indicating the O atoms, closely bound to Co, do not follow easily the fast Na$^+$ motion. Similar effects have been observed in magnetoresistive manganites, where the crossover from adiabatic to non-adiabatic Mn-O bond relaxation across the paramagnetic-insulator to ferromagnetic-metal transition also gives rise to a decrease of the mean-square displacements \cite{Medarde_99}. A dynamic origin would also explain the dramatic changes of the $^{59}$Co relaxation rate at $T_1$ reported by NMR and $\mu^{+}$SR studies, whose unusual temperature dependence may simply reflect the state of the Na layers.\cite{Gavilano_04,Weller_09}

In summary, our data confirm the existence of structural changes in Na$_{0.7}$CoO$_{2}$ at $T_{1}=290$~K and reveal another lattice anomaly at $T_{1}=400$~K. A careful analysis of the crystal structure, the anisotropic Debye-Waller factors and the residual scattering density strongly suggests that both anomalies reflect a response of the lattice to the increasing Na mobility. The structural modifications at $T_1$ shorten 2/3 of the Na1-Na2 distances, creating Q1D pathways which allow for Na diffusion at temperatures as low as 290~K. A full 2D diffusion is only observed above 400 K, coinciding with the stabilization of equal Na-Na distances in the hexagonal structure. The "melting" within the Na layers hence occurs in two steps and involves a crossover from Q1D to 2D Na diffusion. This clarifies two previously incompatible points of view, which associated the structural anomaly at $T_{1}$ either to modifications of the static Na-vacancy clustering or to the 2D-melting of the Na layers. It also shows that tailored ionic diffusion may be used to modify the crystal structure and vice versa. Because of the impact of changes in interatomic distances and angles in the orbital overlap/bandwidth, such insight opens the door to the tuning of physical properties $e.g.$ magnetism or electronic structure, and at the same time increases the possibilities for chemical design of novel materials with improved functional properties.
\bibliographystyle{apsrev}
\vspace{-7mm}
\bibliography{NaxCoO2_LZ13189_revised}

\begin{thebibliography}{22}
\expandafter\ifx\csname natexlab\endcsname\relax\def\natexlab#1{#1}\fi
\expandafter\ifx\csname bibnamefont\endcsname\relax
  \def\bibnamefont#1{#1}\fi
\expandafter\ifx\csname bibfnamefont\endcsname\relax
  \def\bibfnamefont#1{#1}\fi
\expandafter\ifx\csname citenamefont\endcsname\relax
  \def\citenamefont#1{#1}\fi
\expandafter\ifx\csname url\endcsname\relax
  \def\url#1{\texttt{#1}}\fi
\expandafter\ifx\csname urlprefix\endcsname\relax\def\urlprefix{URL }\fi
\providecommand{\bibinfo}[2]{#2}
\providecommand{\eprint}[2][]{\url{#2}}

\bibitem[{\citenamefont{{P.~W. Anderson}}(1973)}]{Anderson_73}
\bibinfo{author}{\bibnamefont{{P.~W. Anderson}}}, \bibinfo{journal}{Mater. Res.
  Bull.} \textbf{\bibinfo{volume}{8}}, \bibinfo{pages}{153}
  (\bibinfo{year}{1973}).

\bibitem[{\citenamefont{{J.~G. Bednorz and
  K.~A.~M\"{u}ller}}(1986)}]{Bednorz_86}
\bibinfo{author}{\bibnamefont{{J.~G. Bednorz and K.~A.~M\"{u}ller}}},
  \bibinfo{journal}{Z. Phys. B - Cond. Mat.} \textbf{\bibinfo{volume}{64}},
  \bibinfo{pages}{189} (\bibinfo{year}{1986}).

\bibitem[{\citenamefont{{D. Linden \& T.~B. Reddy (ed)}}(2010)}]{Linden_10}
\bibinfo{author}{\bibnamefont{{D. Linden \& T.~B. Reddy (ed)}}},
  \emph{\bibinfo{title}{Linden's Handbook of Batteries, 4th Edition}}
  (\bibinfo{publisher}{McGraw-Hill Inc., New York}, \bibinfo{year}{2010}).

\bibitem[{\citenamefont{{K. Takada $et~al.$}}(2003)}]{Takada_03}
\bibinfo{author}{\bibnamefont{{K. Takada $et~al.$}}}, \bibinfo{journal}{Nature}
  \textbf{\bibinfo{volume}{422}}, \bibinfo{pages}{53} (\bibinfo{year}{2003}).

\bibitem[{\citenamefont{{J. Sugiyama $et~al.$}}(204)}]{Sugiyama_04}
\bibinfo{author}{\bibnamefont{{J. Sugiyama $et~al.$}}}, \bibinfo{journal}{Phys.
  Rev. Lett.} \textbf{\bibinfo{volume}{92}}, \bibinfo{pages}{017602}
  (\bibinfo{year}{204}).

\bibitem[{\citenamefont{{L. Balicas $et~al.$}}(2008)}]{Balicas_08}
\bibinfo{author}{\bibnamefont{{L. Balicas $et~al.$}}}, \bibinfo{journal}{Phys.
  Rev. Lett.} \textbf{\bibinfo{volume}{100}}, \bibinfo{pages}{126405}
  (\bibinfo{year}{2008}).

\bibitem[{\citenamefont{{C. Delmas $et~al.$}}(1981)}]{Delmas_81}
\bibinfo{author}{\bibnamefont{{C. Delmas $et~al.$}}}, \bibinfo{journal}{Sol.
  St. Ionics} \textbf{\bibinfo{volume}{3/4}}, \bibinfo{pages}{165}
  (\bibinfo{year}{1981}).

\bibitem[{\citenamefont{{M. Roger $et~al.$}}(2007)}]{Roger_07}
\bibinfo{author}{\bibnamefont{{M. Roger $et~al.$}}}, \bibinfo{journal}{Nature}
  \textbf{\bibinfo{volume}{445}}, \bibinfo{pages}{631} (\bibinfo{year}{2007}).

\bibitem[{\citenamefont{{M.-H. Julien $et~al.$}}(2008)}]{Julien_08}
\bibinfo{author}{\bibnamefont{{M.-H. Julien $et~al.$}}},
  \bibinfo{journal}{Phys. Rev. Lett.} \textbf{\bibinfo{volume}{100}},
  \bibinfo{pages}{096405} (\bibinfo{year}{2008}).

\bibitem[{\citenamefont{{T.~F. Schulze $et~al.$}}(2008)}]{Schulze_09}
\bibinfo{author}{\bibnamefont{{T.~F. Schulze $et~al.$}}},
  \bibinfo{journal}{Phys. Rev. Lett.} \textbf{\bibinfo{volume}{100}},
  \bibinfo{pages}{026407} (\bibinfo{year}{2008}).

\bibitem[{\citenamefont{{M. Weller $et~al.$}}(2009)}]{Weller_09}
\bibinfo{author}{\bibnamefont{{M. Weller $et~al.$}}}, \bibinfo{journal}{Phys.
  Rev. Lett.} \textbf{\bibinfo{volume}{102}}, \bibinfo{pages}{056401}
  (\bibinfo{year}{2009}).

\bibitem[{\citenamefont{{G.~J. Shu $et~al.$}}(2008)}]{Shu_08}
\bibinfo{author}{\bibnamefont{{G.~J. Shu $et~al.$}}}, \bibinfo{journal}{Phys.
  Rev. B} \textbf{\bibinfo{volume}{78}}, \bibinfo{pages}{052101}
  (\bibinfo{year}{2008}).

\bibitem[{\citenamefont{{I.~R. Mukhamedshin $et~al.$}}(2004)}]{Mukhamedshin_04}
\bibinfo{author}{\bibnamefont{{I.~R. Mukhamedshin $et~al.$}}},
  \bibinfo{journal}{Phys. Rev. Lett.} \textbf{\bibinfo{volume}{93}},
  \bibinfo{pages}{167601} (\bibinfo{year}{2004}).

\bibitem[{\citenamefont{{D.~J.~P. Morris $et~al.$}}(2009)}]{Morris_09}
\bibinfo{author}{\bibnamefont{{D.~J.~P. Morris $et~al.$}}},
  \bibinfo{journal}{Phys. Rev. B} \textbf{\bibinfo{volume}{79}},
  \bibinfo{pages}{100103(R)} (\bibinfo{year}{2009}).

\bibitem[{\citenamefont{{F.-T. Huang $et~al.$}}(2009)}]{FTHuang_09}
\bibinfo{author}{\bibnamefont{{F.-T. Huang $et~al.$}}}, \bibinfo{journal}{Phys.
  Rev. B} \textbf{\bibinfo{volume}{79}}, \bibinfo{pages}{014413}
  (\bibinfo{year}{2009}).

\bibitem[{\citenamefont{{Y. Hinuma $et~al.$}}(2008)}]{Hinuma_08}
\bibinfo{author}{\bibnamefont{{Y. Hinuma $et~al.$}}}, \bibinfo{journal}{Phys.
  Rev. B} \textbf{\bibinfo{volume}{77}}, \bibinfo{pages}{224111}
  (\bibinfo{year}{2008}).

\bibitem[{\citenamefont{{Q. Huang $et~al.$}}(2004)}]{QHuang_04}
\bibinfo{author}{\bibnamefont{{Q. Huang $et~al.$}}}, \bibinfo{journal}{Phys.
  Rev. B} \textbf{\bibinfo{volume}{70}}, \bibinfo{pages}{184110}
  (\bibinfo{year}{2004}).

\bibitem[{\citenamefont{{J.~L. Gavilano $et~al.$}}(2004)}]{Gavilano_04}
\bibinfo{author}{\bibnamefont{{J.~L. Gavilano $et~al.$}}},
  \bibinfo{journal}{Phys. Rev. B} \textbf{\bibinfo{volume}{69}},
  \bibinfo{pages}{100404(R)} (\bibinfo{year}{2004}).

\bibitem[{\citenamefont{{P. Fischer $et~al.$}}(2000)}]{HRPT}
\bibinfo{author}{\bibnamefont{{P. Fischer $et~al.$}}},
  \bibinfo{journal}{Physica B} \textbf{\bibinfo{volume}{276-278}},
  \bibinfo{pages}{146} (\bibinfo{year}{2000}).

\bibitem[{\citenamefont{{J. Rodr\'{\i}guez-Carvajal}}(1993)}]{Fullprof}
\bibinfo{author}{\bibnamefont{{J. Rodr\'{\i}guez-Carvajal}}},
  \bibinfo{journal}{Physica B} \textbf{\bibinfo{volume}{192}},
  \bibinfo{pages}{55} (\bibinfo{year}{1993}).

\bibitem[{\citenamefont{{H.~W. Zandbergen $et~al.$}}(2004)}]{Zandbergen_04}
\bibinfo{author}{\bibnamefont{{H.~W. Zandbergen $et~al.$}}},
  \bibinfo{journal}{Phys. Rev. B} \textbf{\bibinfo{volume}{70}},
  \bibinfo{pages}{024101} (\bibinfo{year}{2004}).

\bibitem[{\citenamefont{{M. Medarde $et~al.$}}(1999)}]{Medarde_99}
\bibinfo{author}{\bibnamefont{{M. Medarde $et~al.$}}}, \bibinfo{journal}{Phys.
  Rev. Lett.} \textbf{\bibinfo{volume}{83}}, \bibinfo{pages}{1223}
  (\bibinfo{year}{1999}).

\end{thebibliography}
\end{document}